\documentclass[conference]{IEEEtran}
\IEEEoverridecommandlockouts

\usepackage{cite}
\usepackage{amsmath,amssymb,amsfonts}
\usepackage{algorithm}
\usepackage{algorithmic}
\usepackage{graphicx}
\usepackage{subfigure}
\usepackage{textcomp}
\usepackage{xcolor}
\usepackage{bm}
\def\BibTeX{{\rm B\kern-.05em{\sc i\kern-.025em b}\kern-.08em
    T\kern-.1667em\lower.7ex\hbox{E}\kern-.125emX}}

\begin{document}
\title
{
	Learning to Denoise and Decode: A Novel Residual Neural Network Decoder for Polar Codes
}
\author{Zhiwei~Cao, Hongfei~Zhu, Yuping~Zhao, Dou~Li \\ 
	School of Electronics Engineering and Computer Science\\
	Peking University, Beijing, 100871, China\\
	Email:\{cao\_zhiwei, zhuhongfei, yuping.zhao, lidou\}@pku.edu.cn
	\thanks{This work is financially supported by The National Key Research and Development Program of China under the Grant No. 2018YFB1801403.}}
\maketitle

\begin{abstract}
Polar codes have been adopted as the control channel coding scheme in the fifth generation new radio (5G NR) standard due to its capacity-achievable property. Traditional polar decoding algorithms such as successive cancellation (SC) suffer from high latency problem because of their sequential decoding nature.  Neural network decoder (NND) has been proved to be a candidate for polar decoder since it is capable of \textit{one-shot} decoding and parallel computing. Whereas, the bit-error-rate (BER) performance of NND is still inferior to that of SC algorithm.  In this paper, we propose a residual neural network decoder (RNND) for polar codes. Different from previous works which directly use neural network for decoding  symbols received from the channel, the proposed RNND introduces a denoising module based on residual learning before NND. The proposed residual learning denoiser is able to remove remarkable amount of noise from received signals. Numerical results show that our proposed RNND outperforms traditional NND with regard to the BER performance under comparable latency.
\end{abstract}

\section{Introduction}
	Polar codes proposed by Erdal Arikan are the first provable capacity achieving codes for symmetric binary-input discrete memoryless channels (B-DMCs) \cite{Polar-Arikan}. Although successive cancellation (SC) decoding algorithm proposed by Arikan initially has a lower complexity $O(NlogN)$ in terms of the code length $N$, its performance for finite block lengths is not satisfactory. Later successive cancellation list  \cite{SCL-Tal_Vardy,SCL-Niu}, successive cancellation stack  \cite{SCS} and CRC-aided SCL/SCS  decoders \cite{CRC-SCL_SCS} were introduced to improve the decoding performance of polar codes. However, they still suffer from high latency as well as limited throughput due to their sequential decoding property, which is disadvantageous to the practical wireless communication systems requiring high reliability and low latency. 

	Recently deep learning (DL) has attracted worldwide attentions because of its powerful capabilities to solve complicated tasks. With the help of deep learning, significant improvements have been achieved in many fields, such as computer vision\cite{computer_vision}, natural language processing\cite{natural_language_processing}, autonomous vehicles\cite{autonomous_vehicles} and many other areas. In the field of communication, the general channel decoding problems can also be solved with deep learning, since they can be regarded as a type of classification problem. The channel decoder based on deep neural network is called the neural network decoder (NND) consisting of two stages: NND training and NND testing. Non-iterative and consequently low-latency decoding are two main advantages of NND, because NND calculates the  estimated value of information bits by passing each layer only once with the pre-trained neural network, which is referred to as \textit{one-shot} decoding. What's more, the current deep learning platforms, such as Pytorch \cite{pytorch}, and the powerful hardwares like graphical processing units (GPUs) can enable the efficient implementation of NND.

	NND for polar codes was proposed in \cite{NND_initial}, where the authors found that structured codes are indeed easier to learn than random codes, and the bit-error-rate (BER) performance of NND is close to that of maximum a posteriori (MAP) decoding. In \cite{feed-forward} the authors considered a deep feed-forward neural network for polar codes and investigated its decoding performances with respect to numerous configurations: the number of hidden layers, the number of nodes for each layer, and activation functions. Later, more discussion on the activation function of the neural network for decoding polar codes and decoding under Reighlay fading channels using NND can be found in \cite{activation} and \cite{rayleigh}, respectively.  In \cite{ICC}, the authors compared the decoding performance of three types of neural network, i.e., multi-layer perceptron (MLP), convolutional neural network (CNN) and recurrent neural network (RNN) with the same parameter magnitude. The authors found that the neural network is capable of learning the complete encoding structure with noiseless codewords as training data. The comparison of the three types of neural network was also discussed in \cite{polar-LDPC}, where the authors proposed a unified polar-LDPC NND by concatenating an indicator section. 
	
	All results shown in the above papers suggest that higher test-SNR (Signal-to-Noise Ratio) during the NND testing stage facilitates decoding with NND. Dramatically thinking, if we can place a denoising neural network before NND to improve the test-SNR, the modified NND will certainly show a better decoding performance than the original NND. It is well known that residual learning has been widely used in image denoising \cite{resnet-image_denoising}. Thus, it may shed some lights on the codeword denoising.
	
	 Motivated by this, in this paper we propose a residual neural network decoder (RNND) for polar codes. The proposed RNND introduces a denoiser based on residual learning before NND to improve the SNR of received symbols. Simulation results demonstrate that the proposed RNND has a considerable advantage over existing NNDs w.r.t. bit-error-rate (BER) performance under low latency.  

	The rest of this paper is organized as follows. Section \ref{Preliminary} provides some preliminary knowledge of polar codes, three types of neural network (MLP, CNN, RNN) and the residual learning. The system model is introduced in Section \ref{design}. The architecture of the proposed RNND  with a novel multi-task learning objective and the training  and testing strategy are discussed in Section \ref{RNND}. We provide the experiment configuration and corresponding numerical results in Section \ref{Result}.  Eventually, concluding remarks are given in Section \ref{conclution}.

\section{Preliminaries}\label{Preliminary}
\subsection{Polar Codes}
	To construct an $(N,K)$ polar codes, the $K$ information bits $u_\mathcal{A}$ and the other $N-K$ frozen bits $u_{\mathcal{A}^c}$ are first assigned to the reliable and unreliable positions of the $N$-bit message $u_1^{N}$, respectively. $\mathcal{A}$ is the information set, while $\mathcal{A}^c$ is the frozen set which is the complementary set of $\mathcal{A}$. The $N-K$ frozen bits with indices in $\mathcal{A}^c$ are always fixed to zeros. Then the $N$-bit transmitted codeword $x_1^{N}$ can be obtained by multiplying $u_1^{N}$ with the generator matrix $G_N$ as follows
	
	\begin{equation}
		x_1^{N}=u_1^{N}G_N=u_1^{N}B_NF_2^{\otimes n}, n=log_2{N}
	\end{equation}

	where $ \otimes $ denotes the Kronecker product, $ {F}_2 = \left[ {\begin{array}{*{20}{c}} 
	1&0\\
	1&1
	\end{array}} \right] $ and $ {B}_N $ is the bit-reversal permutation matrix.\par

\subsection{Neural Network}
In this paper, we utilize three popular and effective neural network architectures: 1) MLP; 2) CNN; 3) RNN to construct RNNDs and compare their BER performances. Here, we briefly describe the general architecture of these neural networks.
\subsubsection{ MLP}
\begin{figure}[t!]
	\centering
	\includegraphics[width=.45\textwidth]{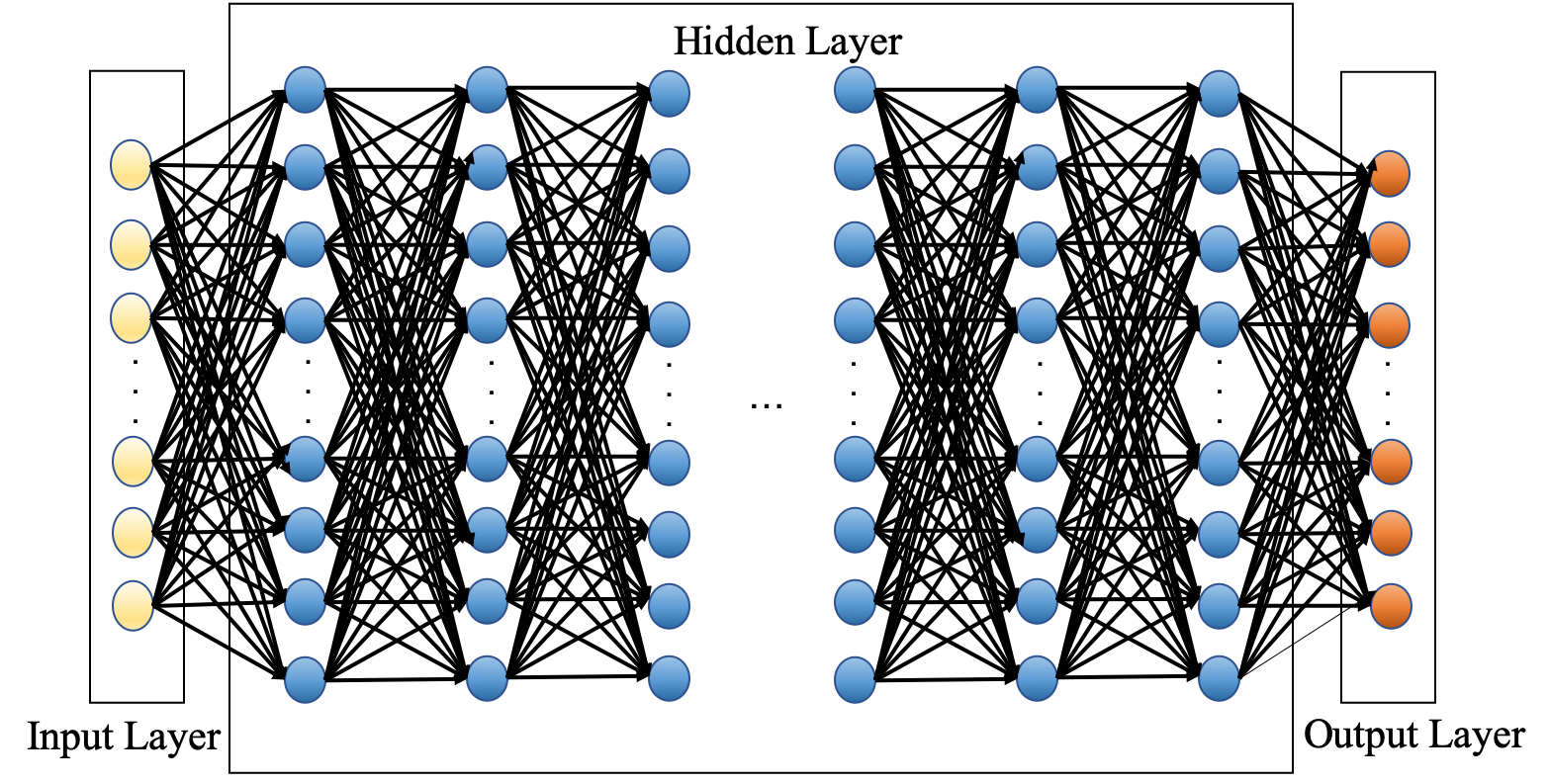}
	\caption{Diagram of multi-layer perceptron (MLP).} 
	\label{DNN} 
\end{figure}

MLP is a kind of feedforward and densely connected neural network. There is no feedback or loop in the neural network and each node is connected with all its ancestors. The input data is processed layer by layer with affine transformation and nonlinear activation function between layers and finally reaches the output layer.  Fig. \ref{DNN} illustrates the general architecture of MLP.
\subsubsection{CNN}
\begin{figure}[t!]
	\centering
	\includegraphics[width=.45\textwidth]{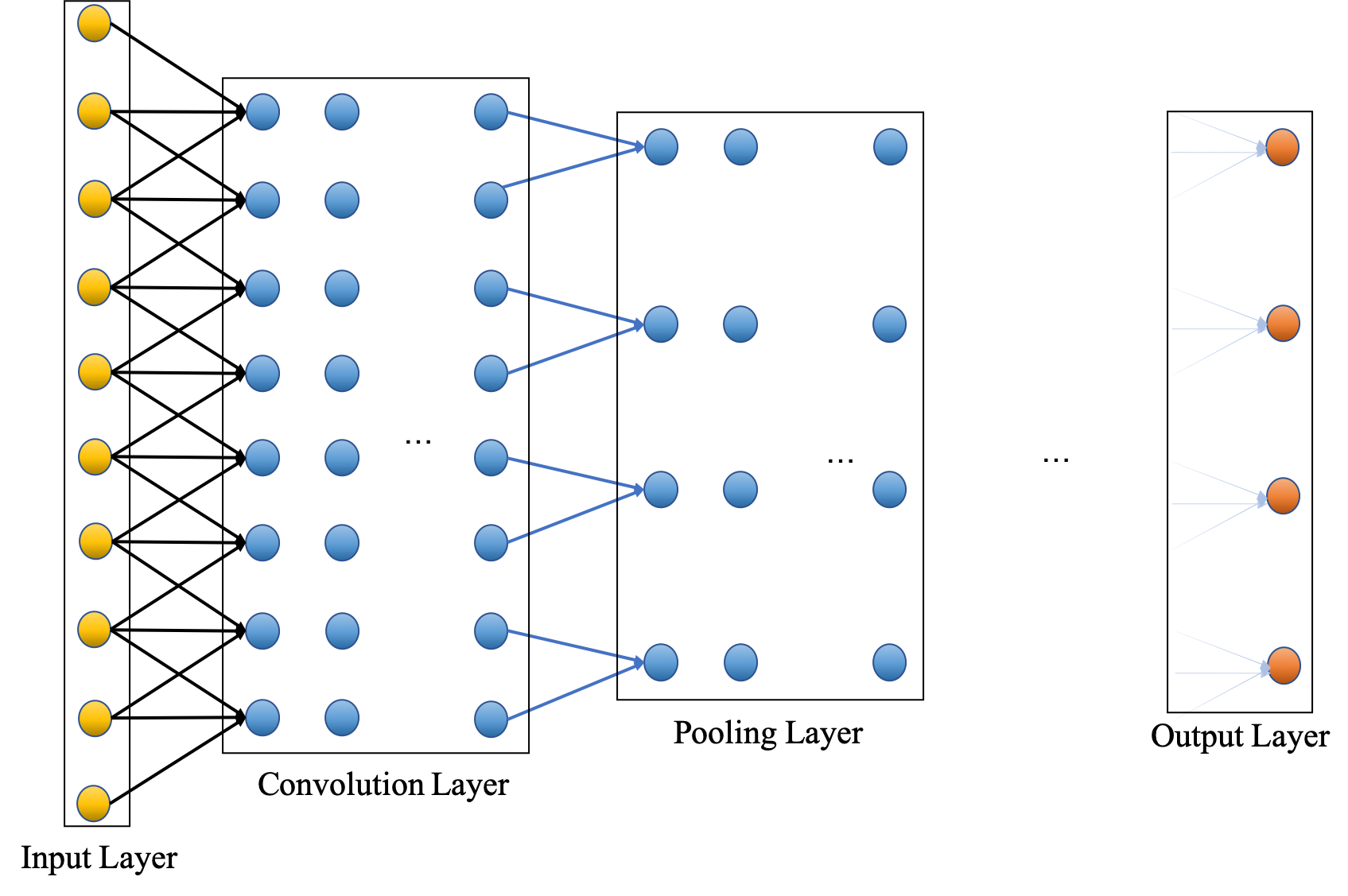}
	\caption{Diagram of convolutional neural network (CNN). The black and blue arrows represent affine transformation and pooling operators, respectively. Better view in color.} 
	\label{CNN} 
\end{figure}

The hidden layers of classical CNN are either convolutional layers or pooling layers. It is universally known that CNN has strong ability of extracting features automatically. Fig. \ref{CNN} shows the classical diagram for CNN. 

\subsubsection{RNN}
\begin{figure}[t!]
	\centering
	\includegraphics[width=.5\textwidth]{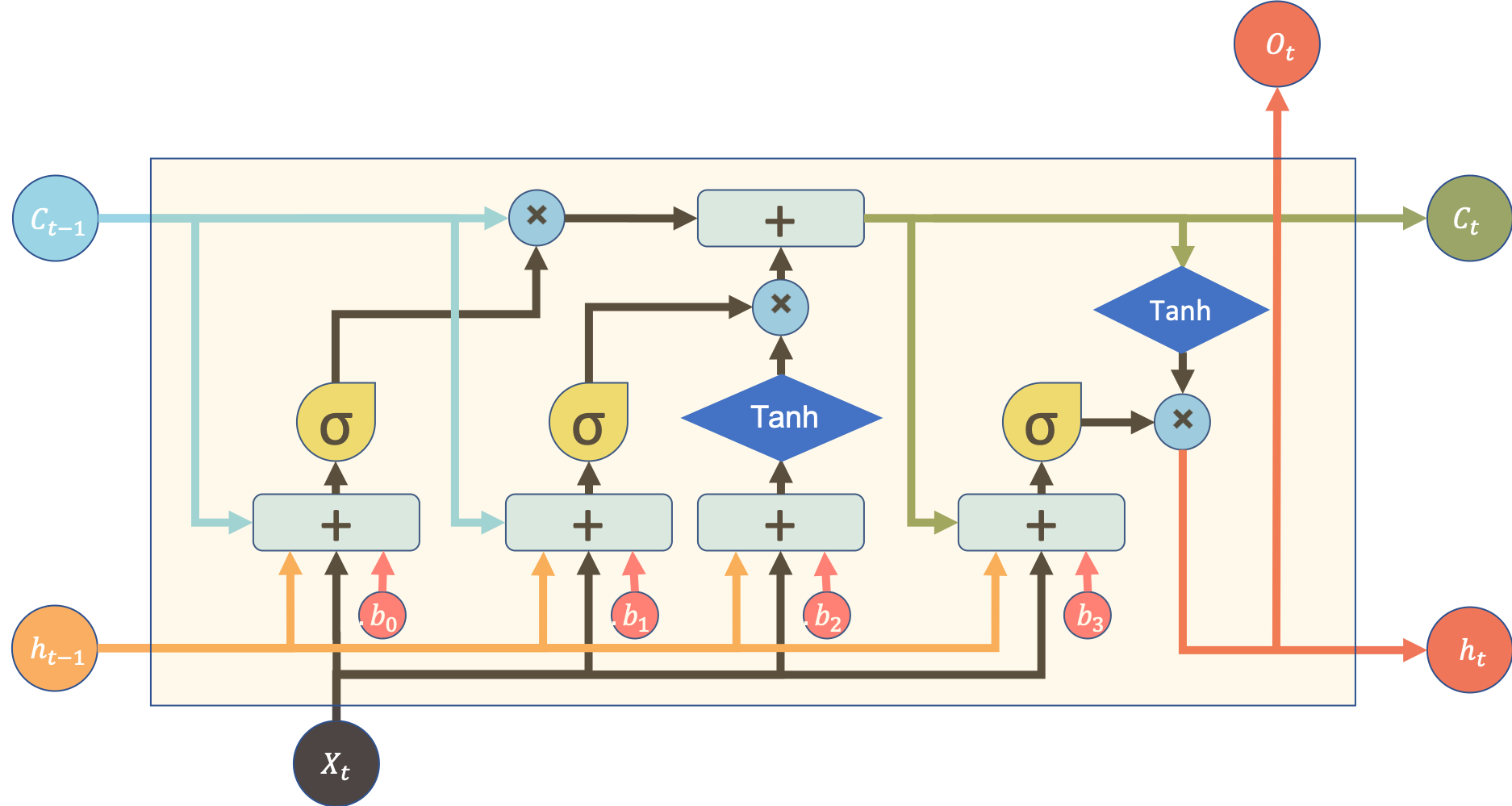}
	\caption{Diagram of long short-time memory (LSTM). $h_t, C_t, X_t$ and $O_t$ represent the hidden vector, cell state, input vector and output vector at time $t$, respectively.} 
	\label{LSTM} 
\end{figure}

RNN is a class of neural network that has recurrent structure, i.e., previous states will influence current outputs. Therefore, RNN is very powerful  for time series modeling. Traditional RNN suffers from severe vanishing and exploding gradient problem\cite{vanish}, making it difficult to train. Hence, in practice, people usually utilize its variants like LSTM\cite{LSTM} and GRU\cite{GRU}. In this paper, we use LSTM as our practical RNN implementation.  Fig. \ref{LSTM}  presents the classical structure of the LSTM. 

\subsection{Residual Learing}
\begin{figure}[t!]
	\centering
	\includegraphics[width=.3\textwidth]{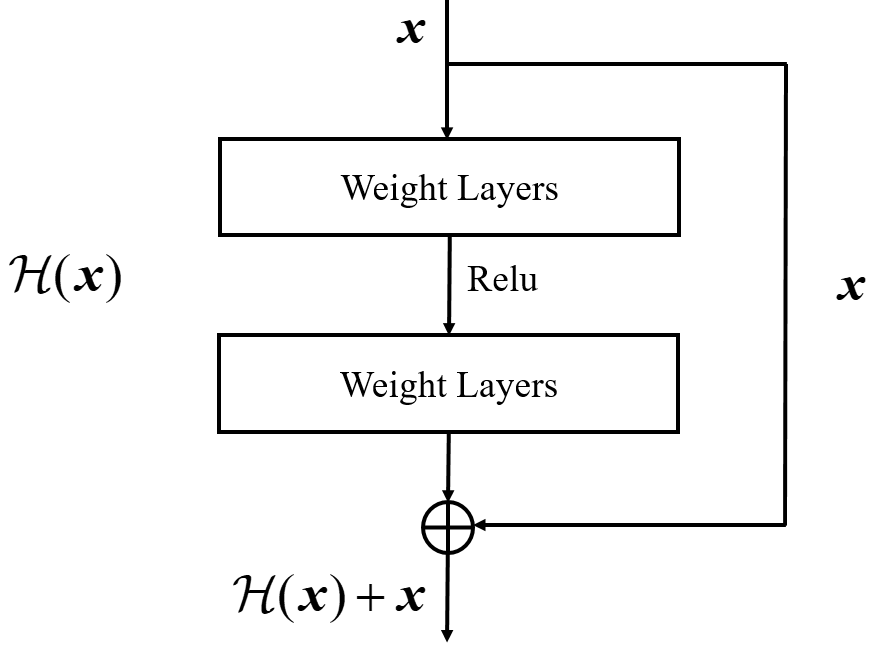}
	\caption{Diagram of the residual learning block. $\mathcal{H}(\bm{x})$ corresponds to the stacked weight layers with Relu non-linearity. } 
	\label{ResBlock} 
\end{figure}

Residual learning of deep neural network was first proposed in \cite{ResNet} to solve the performance degradation problem, i.e., the training accuracy begins to decrease as the depth of the neural network increases. By assuming the residual mapping is much easier to learn, the authors in \cite{ResNet} stacked a few nonlinear layers for explicitly learning the residual mapping. Residual learning of neural networks is achieved through shortcut connections. Fig. \ref{ResBlock} illustrates the general architecture of the residual learning block.

\section{System  Model}\label{design}

\begin{figure*}[t!]
	\centering
	\includegraphics[width=.95\textwidth]{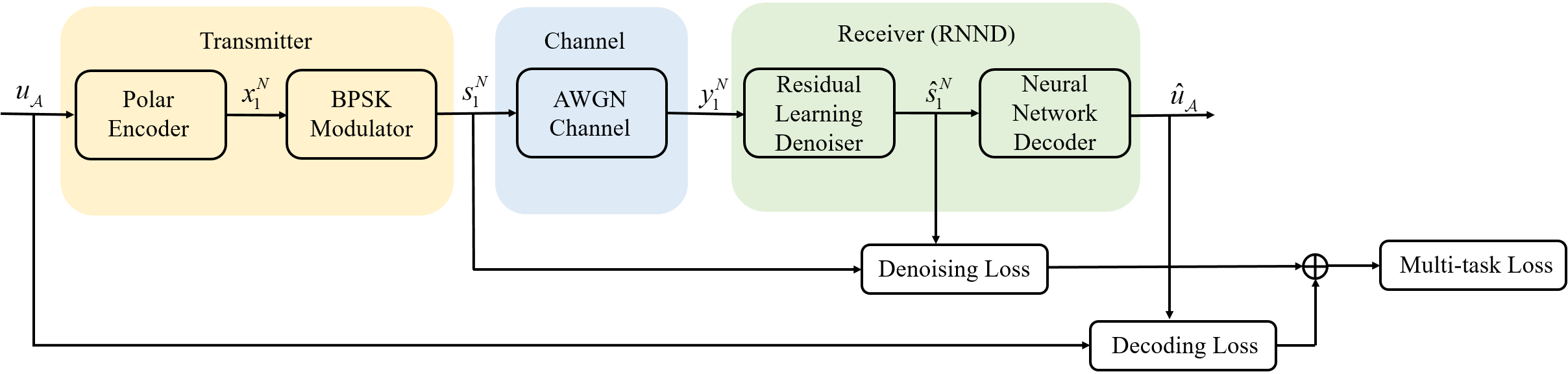}
	\caption{The system model in this paper. Residual learning denoiser and neural network decoder make up the architecture of  the proposed RNND. The denoising loss and decoding loss correspond to (\ref{loss_denoise}) and (\ref{loss_decode}), respectively. The multi-task loss is the proposed joint optimization objective (\ref{loss}).} 
	\label{Model} 
\end{figure*}

The system model  in this paper is shown in Fig. \ref{Model}. At the transmitter,  $K$-bit information $u_\mathcal{A}$ is first encoded into a $N$-bit codeword $x_1^N$ with a polar encoder. Then $x_1^N$ is mapped to $N$ modulated symbols $s_1^N$ through binary phase shift keying (BPSK) modulation by $s_i = 1 - 2x_i$ for $i=1,2,...,N$. The modulated symbol $s_1^N$ is subsequently trainsmitted over an additive white Gaussian noise (AWGN) channel. 

The signal model for received symbols can be formulated as follows
\begin{align}
y_1^N = s_1^N + n_1^N
\end{align}
where $n_1^N \sim \mathcal{N}(\bm{0}, \sigma^2\mathbf{I}_N)$ represents the i.i.d. Gaussian noise vector. 

At the receiver, the received symbol vector $y_1^N$ first enters the residual learning denoiser to reduce the noise. After the denoising process, the filtered signal vector $\hat{s}_1^N$ is then sent to the neural network decoder. Finally, we obtain the estimated value of $u_{\mathcal{A}}$, denoted as $\hat{u}_{\mathcal{A}}$. The two modules, residual learning denoiser and neural network decoder, make up the architecture of  the proposed RNND which will be addressed in Section \ref{RNND}. Addtionally, the three loss function calculation modules which are denoising loss, decoding loss and multi-task loss shown in Fig. \ref{Model} will be discussed in detail in Section \ref{RNND}.

\section{Proposed Residual Neural Network Decoder}\label{RNND}
In this section, we first delineate the architecture of the proposed RNND which contains the residual learning denoiser and the neural network decoder. The denoiser and decoder will be jointly trained through a novel multi-task learning objective which will be discussed in the following. Ultimately, we depict the  training and testing strategy for jointly learning the decoder and the denoiser.  
\subsection{Residual Learning Denoiser}
 Residual learning has been shown to be a powerful tool for image denoising\cite{resnet-image_denoising} and  the codeword denoising can be regarded as the special case of image denoising in one-dimension input. Residual learning module, i.e., $\mathcal{H}(\bm{x})$ in Fig. \ref{ResBlock}, can be directly viewed as a denoising module as we optimize its parameters, so that the output of it, i.e., $\mathcal{H}(\bm{x})+\bm{x}$ in Fig. \ref{ResBlock}, can be as close as possible to the transmitted symbols. The residual learning denoiser consists of some stacked layers and a shortcut connection, as shown in Fig. \ref{ResBlock}. The weight layers in Fig. \ref{ResBlock} can be any type of neural network. In this work, we investigate three different types  of neural network as the denoiser, which are MLP, CNN and RNN, respectively. 
 
 As shown in Fig. \ref{Model}, denote the function corresponding to the stacked weight layers in the denoiser as $\mathcal{H}(y_1^N)$ and the output of the denoiser is the element-wise summation over the received signal vector $y_1^N$ and $\mathcal{H}(y_1^N)$
\begin{align}
\hat{s}_1^N = y_1^N + \mathcal{H}(y_1^N)
\end{align}
For the denoising task, we aim to minimize the difference of $\hat{s}_1^N$ and the transmitted symbol $s_1^N$. We take mean squared error (MSE) to measure the discrepency between $\hat{s}_1^N$ and $s_1^N$ and it leads to our loss function for the denoising task
\begin{align}
\mathcal{L}_{denoise} = \frac{1}{N}||\hat{s}_1^N - s_1^N||_2^2\label{loss_denoise}
\end{align}


\subsection{Neural Network Decoder}
The neural network decoder is illustrated in Fig. \ref{Model}, following the residual learning denoiser. In our work, we regard channel decoding as a  binary classification problem: the decoder attempts to categorize each received symbol into 0 or 1. For the classification problem, there are some commonly used loss functions like binary cross entropy (BCE) and MSE. In \cite{NND_initial} the author has shown via experiments that MSE and BCE have little difference w.r.t. the BER performance of NNDs. Here, we utilize MSE as our loss function for decoding task
\begin{align}\label{loss_decode}
\mathcal{L}_{decode}(\hat{u}_{\mathcal{A}}, u_{\mathcal{A}}) &= \frac{1}{K}||\hat{u}_{\mathcal{A}}- u_{\mathcal{A}}||_2^2\\\nonumber
& = \frac{1}{K}||\mathcal{G}(\hat{s}_1^N) - u_{\mathcal{A}}||_2^2
\end{align}
where, $\mathcal{G}$ is the function corresponds to the neural network decoder.

\subsection{Multi-task Learning}
As mentioned above, our goal is to train the entire neural network so that it can accomplish both denoising and decoding tasks. One potential way for achieving such goal is multi-task learning. In our work, we explicitly sum the decoding loss and denoising loss together as our final loss function 
\begin{align}\label{loss}
\mathcal{L} = \mathcal{L}_{denoise} + \mathcal{L}_{decode}
\end{align} 
With multi-task learning, we hope the learning process of the denoiser and that of the decoder reinforce mutually. The gradient signal from the decoder may lead to better learning of the denoiser. Meanwhile the proposed denoiser will reduce noise level of the received signal, which facilitates the decoder to learn decoding rules. Since $\mathcal{L}_{denoise}$ and $\mathcal{L}_{denoise}$ are both continous and differentiable everywhere, it can be efficiently optimized with some famous optimizers like SGD\cite{SGD} and Adam\cite{ADAM}. 

\subsection{Training  and Testing  Strategy}
\subsubsection{Training data generation}
To avoid \textit{curse of dimensionality}\cite{CurseOfDimensionality}, we only take polar codes with code length $N=16$ and rate $R=1/2$. With such code setting, there are $2^8=256$ distinct codewords with length $N$. We use all of them for training NNDs. During training, these codewords will be modulated with BPSK and contaminated by the AWGN channel. We set the train-SNR = 0 dB as \cite{NND_initial} suggests. 
\subsubsection{Training procedure}
We use batch-based training method. The batch size is set to be 64 in our experiments. At each iteration, we first select a batch of codewords from the codeword set in order. Then we push these codewords into the model and evaluate the loss function with (\ref{loss}). Finally, we calculate gradients with backpropagation and update model parameters with Adam\cite{ADAM} in an end-to-end fashion. Leaning rate is set to be 0.001 and the momentum is 0.99. After we traverse the whole codeword set, we call one training epoch is over. We jointly train our denoiser and decoder with $2^{16}$ epoches.
\subsubsection{Testing procedure}
After the training stages of different RNNDs, we start the NND testing stage. In contrast to the training stage, the modulated symbols are transmitted over the AWGN channel with different noise levels at the test stage. All results presented in this paper are obtained with the trained models during the testing stage. 

\section{Numerical Results}\label{Result}

In this section, we first describe the design of the neural network architecture. We subsequently calculate the SNR of $y_1^N$ and $\hat{s}_1^N$ with different neural network structures as denoiser for verifying their strong denoising capacity. The probability density function (PDF) of $y_1^N$ and $\hat{s}_1^N$ are further calculated to validate the design. Finally, decoding performance and computation time comparison between NNDs, proposed RNNDs and the SC algorithm will be discussed. 

\subsection{Design of Neural Network Architecture}\label{configure}
In this paper, we name the RNNDs with MLP, CNN and RNN architecture MLP-RNND, CNN-RNND, RNN-RNND, respectively. The configuration of each RNND is portrayed in the following.

\subsubsection{MLP-RNND}
 The proposed architecture of MLP has 3 layers for denoiser and each layer contains 128, 64, 32 nodes, respectively. Following the denoiser, there are another 3 layers MLP for the decoder with 128, 64, 32 nodes in each  layer. The input and output size of the entire system are $N$ and $K$ with no doubt. 
\subsubsection{CNN-RNND}
As CNN is widely used in image processing task, most CNN architectures deal with 2-D input data. In our channel decoding task, we modify the CNN architecture and set the input of CNN as 1-D received signal vectors instead of 2-D images. Meanwhile, we also revise the convolutional layer with 1-D convolution instead of classical 2-D convolution. The denoiser consists of 3 convolutional layers with 64, 48, 32 channels, while the decoder contains 3 convolutional layers with 64, 32, 32 channels. We set the  convolutional kernel size to 3. In addition,  MaxPooling\cite{MaxPooling} with kernel size 2 and stride 2 is utilized between the adjacent convolutional layers.

\subsubsection{RNN-RNND}
In order to make RNN suitable for channel decoding, we regard the channel decoding  task as a time series classification problem. The RNN runs through the input vector with one symbol as input and produces an output vector at each timestep. After the RNN reads the entire input vector, it produces a sequence of output vectors. Due to the recurrent nature of RNN, the last vector in the output sequence encodes the input vector. Thus we select it as the feature vector. It will be further processed with one affine transformation layer to produce the denoising or decoding output. We leverage one layer RNN with output dimension 64 for the denoiser and another one layer RNN with output dimension 48 for the decoder. 

 Each RNND has its NND counterpart named MLP-NND, CNN-NND and RNN-NND for performance comparison. For NNDs, we just cancel the shortcut connection and keep the number of layers the same as their counterpart.  All NNDs and RNNDs have similar amount of parameters to avoid the performance difference coming from the difference of parameter number. The total number of  parameters of six types of neural network is described in detail in Table \ref{parameters}. 
 
\begin{table}[h!]
	\caption{Total Number of Parameters of Six Types of Neural Network}
	\begin{center} 
		\begin{tabular}{|c|c|c|c|c|c|c|}
			\hline  
			\textbf{Type of Neural Network} & \textbf{Total Number of Parameters} \\
			\hline
			MLP-NND                         &  27336                              \\
			\hline
			MLP-RNND                        &  25816                              \\
			\hline
			CNN-NND                         &  25576                              \\
			\hline
			CNN-RNND                        &  25256                              \\
			\hline
			RNN-NND                         &  27208                              \\
			\hline
			RNN-RNND                        &  28376                              \\  
			\hline 
		\end{tabular}
		\label{parameters}
	\end{center}
\end{table}

\subsection{Denoising Capacity Comparison }\label{section:denoise}
\begin{figure}[t!]
	\centering
	\includegraphics[width=.4\textwidth]{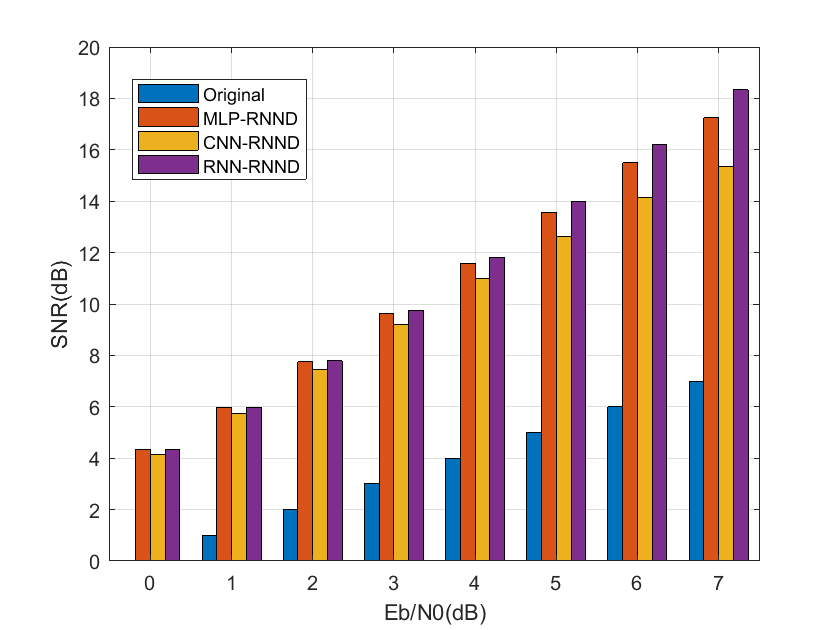}
	\caption{The SNR comparison of signals before and after denoising with different RNNDs. Original refers to the SNR of original received signals.} 
	\label{snr} 
\end{figure}

We investigate the denoising performance of different denoisers in this section. 
Fig. \ref{snr} shows the denoising performance of the denoiser of different neural networks with test-SNR ranging from 0 dB to 7 dB. From Fig. \ref{snr}, we can see the great SNR improvement achieved by the denoisers. For example, MLP-RNND improve the SNR by about 4 dB at test-SNR = dB and more gain can be achieved when the test-SNR increases. It can also be observed that MLP-RNND has similar denoising performance compared with RNN-RNND and both of them substantially outperform the CNN-RNND. It is reasonable to conjecture that the BER performance of MLP-RNND and RNN-RNND will also significantly outperform CNN-RNND since the MLP and RNN denoiser erase more noise than the CNN denoiser. 

\begin{figure}[t!]
	\centering
	\includegraphics[width=.5\textwidth]{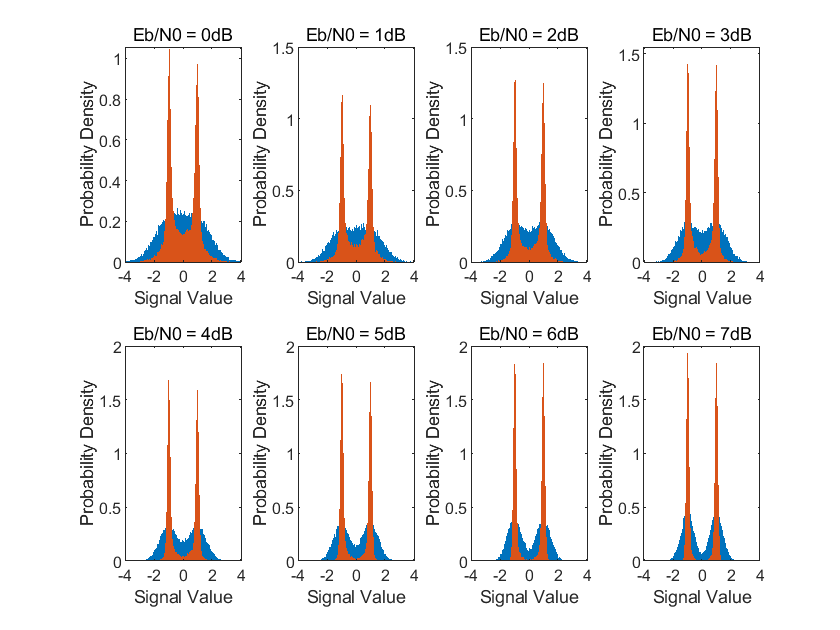}
	\caption{The PDF of signals before and after denoising of the MLP-RNND under different test-Eb/N0. Blue and orange region represent the PDF of received signals and denoised signals, respectively. Better view in color.} 
	\label{pdf} 
\end{figure}

Fig. \ref{pdf} studies the PDF of $y_1^N$ and $\hat{s}_1^N$ of the MLP denoiser under different test-Eb/N0. It is worth noting that the PDF of $\hat{s}_1^N$ is still similar to Gaussian distribution even though the input Gaussian distributed signals $y_1^N$ have gone through nonlinear operations. What's more, it can be found that the variance of the $\hat{s}_1^N$ is much smaller than that of $y_1^N$, which implies considerably amount of Gaussian noise has been removed from received signals. 
\subsection{BER Performance Comparison}
\begin{figure}[t!]
	\centering
	\includegraphics[width=.45\textwidth]{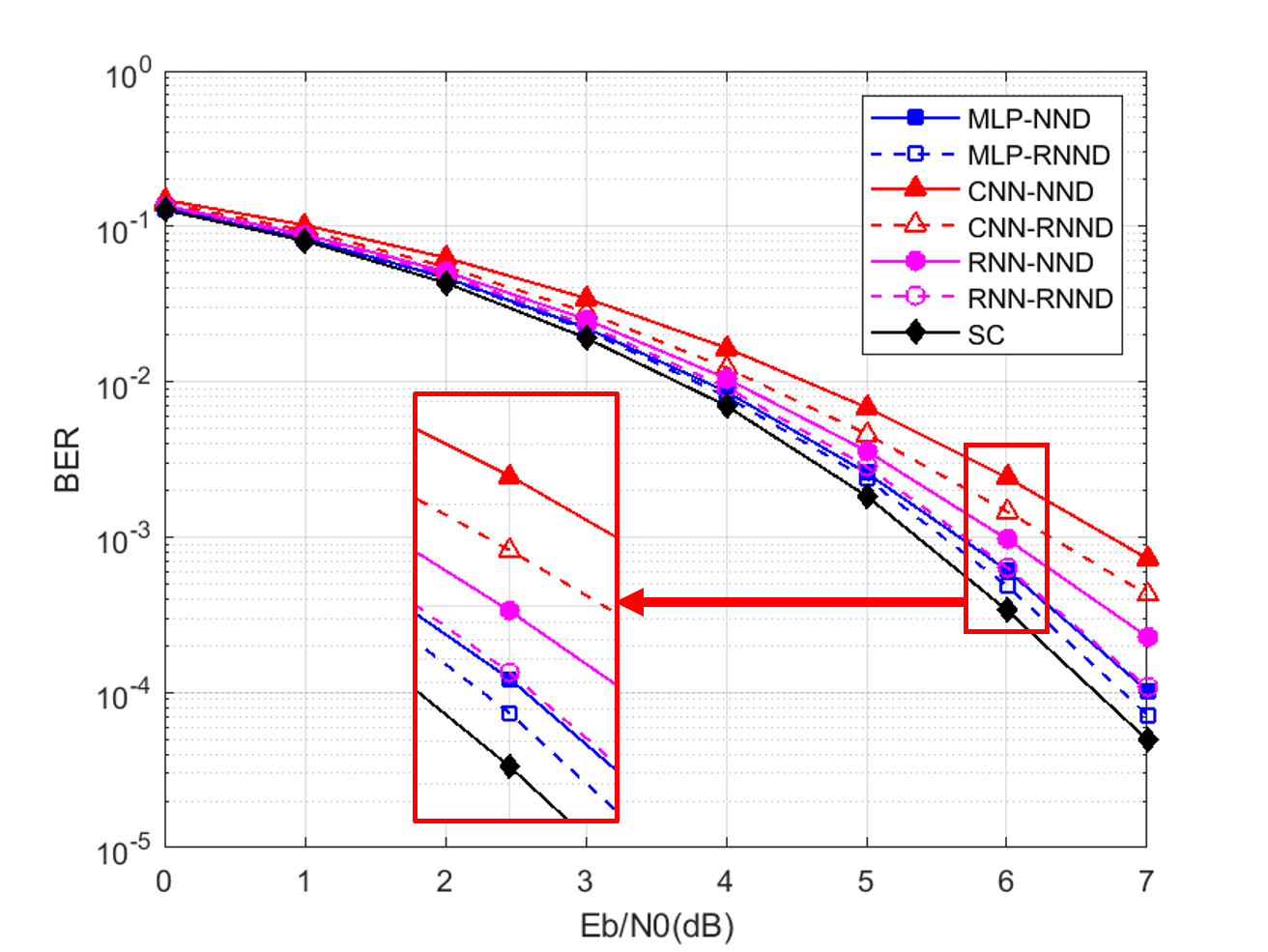}
	\caption{The BER performance comparison of NNDs, proposed RNNDs and the SC algorithm.} 
	\label{ber} 
\end{figure}

After comparing denoising capacity of different denoisers, we study the BER performance of NNDs, proposed RNNDs and the SC algorithm, which is shown in Fig. \ref{ber}.  We can observe that each RNND remarkably outperforms its NND counterpart. For instance, MLP-RNND obtains a gain of roughly 0.2dB over MLP-NND at BER $10^{-4}$. Meanwhile, the BER performance  of MLP-RNND is very close to the SC algorithm which is near optimal in our code setting. It should be pointed out that the BER performances of MLP-RNND and RNN-RNND exceed those of CNN-RNND by a large margin. This is consistent with the conjecture we made in \ref{section:denoise}. It is also important to highlight that MLP-RNND outperforms both CNN-RNND and RNN-RNND, which conforms to previous research\cite{polar-LDPC}. It implies that MLP may be the optimal structure for polar NND.

\subsection{Computation Time Comparison}
\begin{figure}[t!]
	\centering
	\includegraphics[width=.4\textwidth]{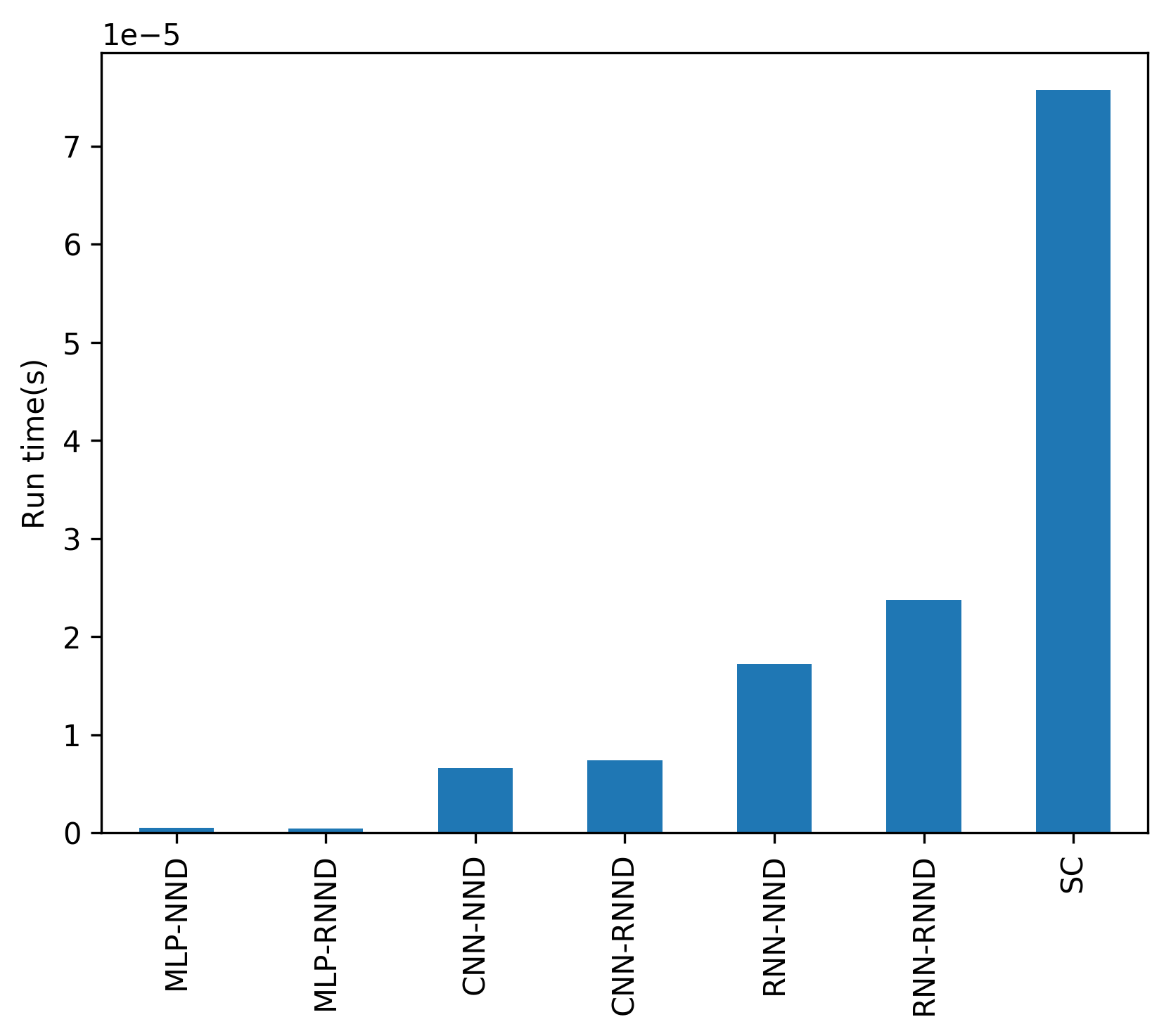}
	\caption{The computation time comparison of NNDs, proposed RNNDs and the SC algorithm.} 
	\label{run-time} 
\end{figure}

We further study the computational time of NNDs, proposed RNNDs and the SC algorithm.  The results are illustrated in Fig. \ref{run-time}. It is worthwhile mentioning that although there exists small gaps between the BER performance of MLP-RNND and that of the traditional SC algorithm, MLP-RNND runs more than 100 times faster than the SC algorithm. It implies that the proposed MLP-RNND may be a strong alternative for traditional SC algorithm and its variants. It must also be mentioned that RNNDs run slightly slower than NNDs whereas such differences are neglectable in practice. 
\section{Conclution}\label{conclution}
In this paper, we propose a novel residual neural network decoder (RNND), which jointly learns to denoise and decode with a multi-task training strategy. We present the PDF and SNR of the signals before and after the trained denoiser, which demonstrates the significant noise level supression of our proposed denoiser. Simulation results show that with the aid of the denoiser, one can obtain considerable BER performance gain since less noisy signals are advantageous to learning decoding rules. The computation time results demonstrate that our proposed MLP-RNND may be a strong competitor with classical SC algorithm due to its near optimal BER performance and  ultra low latency. Notably, although our research focuses on polar codes with short code length, it is also promising to use the proposed RNND to decode polar codes with longer code length since the residual structure facilitates training deep neural network\cite{ResNet}. 

\bibliographystyle{IEEEtran} 
\bibliography{reference}




\end{document}